-----------------------------------------------------------------------
\overfullrule=0pt
\def\refto#1{$^{#1}$}           
\def\ref#1{ref.~#1}                     
\gdef\refis#1{\item{#1.\ }}                    
\def\beginparmode{\endmode
  \begingroup \def\endmode{\par\endgroup}}
\let\endmode=\par
\def\body{\beginparmode}
\def\head#1{                    
  \goodbreak\vskip 0.5truein    
  {\centerline{\bf{#1}}\par}
   \nobreak\vskip 0.25truein\nobreak}
\def\references                 
Phys Rev
  {\head{References}            
   \beginparmode
   \frenchspacing \parindent=0pt \leftskip=1truecm
   \parskip=8pt plus 3pt \everypar{\hangindent=\parindent}}
\def\endreferences{\body}
\def\gs{\mathrel{\raise0.35ex\hbox{$\scriptstyle >$}\kern-0.6em
\lower0.40ex\hbox{{$\scriptstyle \sim$}}}}
\def\ls{\mathrel{\raise0.35ex\hbox{$\scriptstyle <$}\kern-0.6em
\lower0.40ex\hbox{{$\scriptstyle \sim$}}}}
\def\kms{km$\,$s$^{-1}$}

\def\K{{$\,$K}}

\input reforder.tex

\def\mediumspace{\baselineskip 18pt \lineskip 6pt \parskip 3pt plus 5 pt}

\def\today{\number\day\enspace
     \ifcase\month\or January\or Febuary\or March\or April\or May\or
     June\or July\or August\or September\or October\or
     November\or December\fi \enspace\number\year}
\def\clock{\count0=\time \divide\count0 by 60
    \count1=\count0 \multiply\count1 by -60 \advance\count1 by \time
    \number\count0:\ifnum\count1<10{0\number\count1}\else\number\count1\fi}
\footline={\hss -- \folio\ -- \hss}

\def\deg{\ifmmode^\circ\else$^\circ$\fi}
\def\solar{\ifmmode_{\mathord\odot}\else$_{\mathord\odot}$\fi}
\def\jref#1 #2 #3 #4 {{\par\noindent \hangindent=3em \hangafter=1 
      \advance \rightskip by 5em #1, {\it#2}, {\bf#3}, #4.\par}}
\def\ref#1{{\par\noindent \hangindent=3em \hangafter=1 
      \advance \rightskip by 5em #1.\par}}
\newcount\eqnum
\def\nexteq{\global\advance\eqnum by1 \eqno(\number\eqnum)}
\def\lasteq#1{\if)#1[\number\eqnum]\else(\number\eqnum)\fi#1}
\def\preveq#1#2{{\advance\eqnum by-#1
    \if)#2[\number\eqnum]\else(\number\eqnum)\fi}#2}

\def\tableheight{\vrule width 0pt height 8.5pt depth 3.5pt}
{\catcode`|=\active \catcode`&=\active 
    \gdef\tabledelim{\catcode`|=\active \let|=\vbar
                     \catcode`&=\active \let&=\nobar} }
\def\table{\begingroup
    \def\twidth{\hsize}
    \def\tablewidth##1{\def\twidth{##1}}
    \def\defaultheight{\vrule width 0pt height 8.5pt depth 3.5pt}
    \def\heightdepth##1{\dimen0=##1
        \ifdim\dimen0>5pt 
            \divide\dimen0 by 2 \advance\dimen0 by 2.5pt
            \dimen1=\dimen0 \advance\dimen1 by -5pt
            \vrule width 0pt height \the\dimen0  depth \the\dimen1
        \else  \divide\dimen0 by 2
            \vrule width 0pt height \the\dimen0  depth \the\dimen0 \fi}
    \def\spacing##1{\def\defaultheight{\heightdepth{##1}}}
    \def\nextheight##1{\noalign{\gdef\tableheight{\heightdepth{##1}}}}
    \def\end{\cr\noalign{\gdef\tableheight{\defaultheight}}}
    \def\zerowidth##1{\omit\hidewidth ##1 \hidewidth}    
    \def\hline{\noalign{\hrule}}
    \def\skip##1{\noalign{\vskip##1}}
    \def\bskip##1{\noalign{\hbox to \twidth{\vrule height##1 depth 0pt \hfil
        \vrule height##1 depth 0pt}}}
    \def\header##1{\noalign{\hbox to \twidth{\hfil ##1 \unskip\hfil}}}
    \def\bheader##1{\noalign{\hbox to \twidth{\vrule\hfil ##1 
        \unskip\hfil\vrule}}}
    \def\spanloop{\span\omit \advance\mscount by -1}
    \def\extend##1##2{\omit
        \mscount=##1 \multiply\mscount by 2 \advance\mscount by -1
        \loop\ifnum\mscount>1 \spanloop\repeat \ \hfil ##2 \unskip\hfil}
    \def\vbar{&\vrule&}
    \def\nobar{&&}
    \def\hdash##1{ \noalign{ \relax \gdef\tableheight{\heightdepth{0pt}}
        \toks0={} \count0=1 \count1=0 \putout##1\end 
        \toks0=\expandafter{\the\toks0 &\end} \xdef\piggy{\the\toks0} }
        \piggy}
    \let\e=\expandafter
    \def\putspace{\ifnum\count0>1 \advance\count0 by -1
        \toks0=\e\e\e{\the\e\toks0\e&\e\multispan\e{\the\count0}\hfill} 
        \fi \count0=0 }
    \def\putrule{\ifnum\count1>0 \advance\count1 by 1
        \toks0=\e\e\e{\the\e\toks0\e&\e\multispan\e{\the\count1}\leaders\hrule\hfill}
        \fi \count1=0 }
    \def\putout##1{\ifx##1\end \putspace \putrule \let\next=\relax 
        \else \let\next=\putout
            \ifx##1- \advance\count1 by 2 \putspace
            \else    \advance\count0 by 2 \putrule \fi \fi \next}   }
\def\tablespec#1{
    \def\vdimens{\noexpand\tableheight}
    \def\tabby{\tabskip=0pt plus100pt minus100pt}
    \def\r{&################\tabby&\hfil################\unskip}
    \def\c{&################\tabby&\hfil################\unskip\hfil}
    \def\l{&################\tabby&################\unskip\hfil}
    \edef\templ{\noexpand\vdimens ########\unskip  #1 
         \unskip&########\tabskip=0pt&########\cr}
    \tabledelim
    \edef\body##1{ \vbox{
        \tabskip=0pt \offinterlineskip
        \halign to \twidth {\templ ##1}}} }

\newbox\grsign \setbox\grsign=\hbox{$>$}
\newdimen\grdimen \grdimen=\ht\grsign
\newbox\laxbox \newbox\gaxbox
\setbox\gaxbox=\hbox{\raise.5ex\hbox{$>$}\llap
	{\lower.5ex\hbox{$\sim$}}}\ht1=\grdimen\dp1=0pt
\setbox\laxbox=\hbox{\raise.5ex\hbox{$<$}\llap
	{\lower.5ex\hbox{$\sim$}}}\ht2=\grdimen\dp2=0pt

\def\uJy{\ifmmode{\,\mu{\rm Jy}}\else$\,{\mu{\rm Jy}}$\fi}
\def\mJy{\ifmmode{\,{\rm mJy}}\else${\,{\rm mJy}}$\fi}
\def\MHz{\ifmmode{\,{\rm MHz}}\else{$\,{\rm MHz}$}\fi}
\def\GHz{\ifmmode{\,{\rm GHz}}\else{$\,{\rm GHz}$}\fi}
\def\solar{\ifmmode_{\mathord\odot}\else$_{\mathord\odot}$\fi}
\def\Msolar{\ifmmode{\, {\rm M\solar}}\else{${\, {\rm M\solar}}$}\fi}
\def\Rsolar{\ifmmode{\, {\rm R\solar}}\else{${\, {\rm R\solar}}$}\fi}
\def\kms{\ifmmode{\,{\rm km\,s^{-1}}}\else${\,{\rm km\,s^{-1}}}$\fi}
\def\kpc{\ifmmode{\,{\rm kpc}}\else${\,{\rm kpc}}$\fi}
\def\us{\ifmmode{\,\mu{\rm s}}\else$\,{\mu{\rm s}}$\fi}
\def\ms{\ifmmode{\,{\rm ms}}\else$\,{{\rm ms}}$\fi}
\def\y{\ifmmode{\,{\rm y}}\else$\,{\rm y}$\fi}
\def\h{\ifmmode{^{\rm h}}\else$^{\rm h}$\fi}
\def\m{\ifmmode{^{\rm m}}\else$^{\rm m}$\fi}
\def\s{\ifmmode{^{\rm s}}\else$^{\rm s}$\fi}
\def\Lmin{\ifmmode{L_{min}}\else{$L_{min}$}\fi}

\input psfig.sty
\magnification=\magstep1
\mediumspace
\font\eightrm=cmr8
\font\lgh=cmbx10 scaled \magstep2
\def\hb{\hfill\break}



\noindent{\hfill Version: \today, C07037 LS/ks}
\smallskip
\hrule
\bigskip
\line{\lgh A median redshift of 2.4 for galaxies \hb}
\line{\lgh bright at submillimetre wavelengths \hb} 
\bigskip

\line{S.\ C.\ Chapman,$^*$ A.\ W.\ Blain,$^{\star}$ R.\ J.\
  Ivison$^{\dagger}$ and Ian Smail$^{\ddagger}$ \hb}

\bigskip

\line{$^*$
      California Institute of Technology, Pasadena, CA 91125, USA \hfill}

\line{$^\dagger$
      Astronomy Technology Centre, Royal Observatory, Blackford Hill, \hfill}
\line{$\;\,$ Edinburgh EH9 3HJ, UK \hfill}

\line{$^\ddagger$
      Institute for Computational Cosmology, University of Durham,
      South Road \hfill}
\line{$\;\,$ Durham DH1 3LE, UK \hfill}

\bigskip
\hrule
\bigskip

\noindent
{\bf A significant fraction of the energy emitted in the early Universe
came from very luminous galaxies that are largely hidden at optical
wavelengths (because of interstellar dust grains); this energy now forms
part of the cosmic background radiation at wavelengths
near 1$\,$mm.\refto{Fixsen} These submillimetre (submm) galaxies
were resolved from the background in 1997,\refto{SIB} but have been
difficult to identify and study due to the poor spatial resolution of
submm instruments.\refto{Smail02} This has impeded the determination
of their distances (or redshifts, $z$), a crucial element in
understanding their nature and evolution.\refto{BSIK} Here we report
spectroscopic redshifts for ten representative submm galaxies
that we identified reliably using high resolution radio
observations.\refto{BCR,Chapmanradio01,I02} The median redshift for
our sample is 2.4, with a quartile range of $z$ = 1.9--2.8. The submm
population therefore coexists with the peak
activity of quasars, which are thought to be massive black holes in
the process of accreting matter, suggesting a close relationship
between the growth of massive black holes and luminous
dusty galaxies.\refto{Boyle} The space density of
submm galaxies at $z>2$ is about 1000 times greater than that
of similarly luminous galaxies in the present-day Universe, so
they represent an important component of star formation at high
redshifts.}

\bigskip

\vfill\eject

The first real insight in the origin of the far-IR/submm background
came with the commissioning of the SCUBA submm camera on the JCMT.
Almost 100\% of the submm background has been resolved with 
the deepest SCUBA maps, which exploit a sensitivity boost from gravitational 
lenses.\refto{SIB,Smail02,Cowie02}
However,
the near-IR/optical faintness of the submm galaxy population
conspires with the modest positional precision available from the
large ($15''$) SCUBA beam and the large surface density of unrelated
optically-faint galaxies to render positional coincidence alone 
inadequate to identify counterparts.\refto{Smail02,BCR} This
problem is compounded by the small field of view of SCUBA: compiling 
large samples of submm galaxies is slow, with roughly one 
detected every night.  Because of these difficulties, robust
spectroscopic redshifts have been published for only a handful of submm
sources, all atypically bright at optical
wavelengths.\refto{I98,Barger99,I00,Smail02}

Here, we have overcome these problems by taking advantage of 1.4-GHz
radio data from the Very Large Array (VLA) radio telescope
(Hubble Deep Field\refto{Richards00}, 
SSA13 field\refto{Fomalont02},
ElaisN2 and Lockman fields\refto{I02}). These deep
images have fine (1--2$''$) spatial resolution and a large (30$'$
FWHM) field of view. 
The radio data was exploited to followup SCUBA submm data
in the Hubble Deep Field (12\ hr RA) and  SSA13 
(13\ hr RA) fields\refto{Barger98,BCR,Chapmanmega02}, 
and the Lockman (10\ hr RA) and
ElaisN2 (16\ hr RA) fields\refto{SScott02}.
The radio data are sensitive to the synchrotron radio
emission from cosmic-ray electrons accelerated in the supernovae
explosions of the same high-mass stars that heat the dust and generate
the far-IR/submm emission.  Hence a deep radio image
should be an efficient route to pinpoint the positions of many submm
galaxies. This is confirmed in fields with both submm
and radio observations: radio counterparts brighter than 30$\,\mu$Jy
can be found for 60--70\% of submm galaxies brighter than 5$\,$mJy at
850$\,\mu$m (which have a surface density of
450$\,$deg$^{-2}$,\refto{Smail02} and contribute about 20\% of the
submm background).\refto{I02,Chapmanmega02} Moreover, the accurate radio 
positions mean that SCUBA's efficient
`photometry' mode can be used to search for submm
galaxies,\refto{BCR,Chapmanradio01,Chapmanradio02} raising the
detection rate to around ten per night.

Although submm galaxies have faint optical continuum
emission,\refto{Smail02} the radio-detected samples are sufficiently
large to allow efficient, multi-object optical spectroscopy 
using 10-m telescopes.  Slit positions can be assigned
accurately, as the radio/submm emission can be located on optical
images to within 0.3--0.8$''$.\refto{I02} Using this approach we
targetted 34 radio-detected submm galaxies mostly brighter than 5$\,$mJy at
850$\,\mu$m with optical magnitudes in the range $I$ = 22.2--26.4
using the ESI and LRIS spectrographs on the Keck telescopes
(Fig.~1). We measured redshifts for ten submm galaxies, 
representative of the blank-field population. We have doubled the number 
of optically bright ($I<23.5$) submm galaxies 
with accurate redshifts. Our six redshifts for $I>23.5$ submm galaxies
represent the first redshifts for this class of submm galaxy.
The redshifts
span the range $z$ = 0.8--3.7, with a median of 2.4 and an
interquartile range of 1.9--2.8 (Fig.~2). These redshifts allow 
dust temperatures $T_{\rm d}$ and bolometric
luminosities to be determined (see Table~1), assuming
that the tight correlation between far-IR and radio emission observed
at low redshifts\refto{Condon} remains valid.
Typical values are of order 35$\,$K and several
10$^{12}\,$L$_\odot$ respectively.
Note that
relativistic electrons accelerated in shocks close to an active
galactic nucleus (AGN) should boost the radio flux density above that
expected from the standard correlation. If the radio emission is
boosted by an AGN then both temperature and luminosity are
overestimated. There are no obvious cases in the present sample in
this category. 

The redshifts impose robust limits to 
the space densities of submm galaxies in the redshift ranges
0.5--1.2, 1.8--2.8 and 2.8--4 (the gap at $z$ = 1.2--1.8 is the {\it
spectroscopic desert} in which no strong restframe UV lines are
redshifted into the optical window): we calculate $\rho = (3.3 \pm
2.3) \times 10^{-6}$ Mpc$^{-3}$, $(6.5 \pm 2.5) \times
10^{-6}$ Mpc$^{-3}$ and $(2.4 \pm 1.2) \times 10^{-6}$ Mpc$^{-3}$,
respectively, assuming a flat $\Lambda=0.7$ Universe with $H_0$ = 65
km s$^{-1}$ Mpc$^{-1}$.  This compares to a space density of
$\sim10^{-8}$ Mpc$^{-3}$ for low-redshift galaxies with comparable 
bolometric
luminosities greater than $5\times 10^{12}\,$L$_\odot$, comparable to the
luminosities of our spectroscopically targeted galaxies at any
redshift greater than one.  
The volume density of very luminous
galaxies thus increases almost a thousandfold from $z$ $\sim$ 0 to 2.

The relationship between different classes of high-redshift galaxy is
critical for our understanding of galaxy evolution.  Our spectroscopy
shows that the submm galaxies are coeval with the important
populations of star-forming galaxies and quasars detected at $z$ =
2--3 using optical and X-ray
techniques.\refto{Boyle,Barger02,Steidel99,Steidel02} 
We can use their
relative space densities to compare and relate these populations. 
In a 1000~Mpc$^3$ box at $z\sim2.5$, there are
10 Lyman-break galaxies with $R_{\rm AB}$ $<$ 25.5 ($\sim\,$0.2$\times
$L$^\ast_{1500}$) and 1 submm galaxy from our sample (with 850$\mu$m
flux $>$5mJy), but the single submm galaxy produces a comparable
luminosity to the 10 Lyman-break galaxies.
(Here $R_{\rm AB}$ is the flux normalized magnitude near 6900\AA;
$L^\ast_{1500}$ is the characteristic luminosity at restframe 1500\AA,
denoting a change in slope in the luminosity function.)
Equally, the volume density of submm galaxies is similar to
that of narrow-line Type-II AGN selected at these redshifts, from both
X-ray and optical surveys, and an order of magnitude greater than that
of bright optical quasars with $M_B > -25$.\refto{Boyle}  However,
the links between the different classes of sources all depend critically 
on whether the activity is intermittent or continuous.  

To quantify the whole submm population using this 
survey we must account for several selection effects.
In particular, requiring a radio identification prior to
spectroscopy limits the maximum redshift.  Studies of
galaxies at low and moderate redshifts suggest that changes in the
dust properties, especially the dust temperature $T_{\rm d}$, modify
the relative strength of the emission in the submm and radio
wavebands.\refto{B02} The detectability of distant submm galaxies (at a
fixed 850-$\mu$m flux) in the radio waveband thus depends on the
dust temperature: warmer galaxies are detectable at $z$ $\ge$ 3,
while the radio emission from cooler sources falls below the flux
limit at lower redshifts.\refto{Chapmanradio02} About 30\% of
submm galaxies that are brighter than 5$\,$mJy at 850$\mu$m are not
detected in the radio: these galaxies would be lost from our 
catalogue at 
$z \sim 2.5$ for the coldest plausible $T_{\rm d} \sim$ 25\K, and at 
$z>3$ for the typical $T_{\rm d} \simeq 35\,$K that we find. 
Hence, we can only place a lower limit on the space
density in the highest redshift interval. The density derived at
$z=2$ should be unaffected by the need for the galaxies to
be detected at radio wavelengths.

We must also consider the completeness of our spectroscopic
observations. While our redshift completeness is relatively high
($\sim$30\%), we failed to find convincing redshifts for 24 galaxies. 
However, nine of
of these have strong single emission line detections which could plausibly
be identified as Lyman-$\alpha$. Some of the remainder without emission lines
could lie in the spectroscopic desert, making
redshifts difficult to identify. The absence of sources at $z\sim1.5$ 
in Table~1 suggests that this effect
is probably responsible for some of the incompleteness. 
Our targets
generally have faint UV continua, and so there is a bias towards
identifying redshifts for galaxies with strong emission lines. If the
missing galaxies lie outside of the desert, but lack strong emission
lines, then their radio detection suggests they probably lie in the
same $z$ = 1--4 range spanned by the spectroscopically identified
sources.

On the basis of bolometric luminosities estimated from our radio and
submm measurements and redshifts (Table~1), assuming the far-IR--radio
correlation, we calculate the following luminosity densities from our
survey: $\rho_{\rm L}(0.5<z<1.2) = (5 \pm 5)
\times 10^{6}\,$L$_\odot$ Mpc$^{-3}$,
$\rho_{\rm L}(1.8<z<2.8) = (7^{+6}_{-4}) \times
10^{7}\,$L$_\odot$ Mpc$^{-3}$ and $\rho_{\rm L}(2.8<z<4.0) =
(7^{+10}_{-5}) \times 10^{7}\,$L$_\odot$ Mpc$^{-3}$.  To
translate these values into star-formation densities we must subtract
any contribution to the luminosities of these galaxies from the
heating of dust by an active galactic nucleus (AGN).  The proportion
of the submm population detected in deep {\it Chandra} and {\it
XMM-Newton} X-ray surveys, which are sensitive to even dust-obscured
AGN,\refto{I02,Barger02} at flux densities significantly greater than
that expected from star formation alone is at most 30\%. Hence, we assume
conservatively that the far-IR luminosity is derived from star
formation in the remaining 70\%.
Luminosity-density measurements can be translated into a star-formation
density using the standard calibration of 1.6--2.2$\times
$10$^{9}\,$L$_\odot$ (M$_\odot$ yr$^{-1})^{-1}$.\refto{Kennicutt}

The bright radio-detected submm galaxies presented here represent just
20\% of the 850-$\mu$m background, yet their estimated star-formation
densities at redshifts of 2 and 3 are comparable with those inferred
from optical observations.\refto{Steidel99} Both of these estimates
require corrections, in the case of the optical surveys an increase of
a factor of 5 is needed to account for dust
obscuration,\refto{Pettini} while our submm estimate needs to be
corrected by a similar factor to account for spectroscopic
incompleteness and sources below our submm flux limit.  
We find that the submm and Lyman-break
populations encompass all of the star formation activity needed to
reproduce the star formation history inferred from models of the
background radiation and submm counts.\refto{BSIK} Hence, our
substantially complete redshift survey provides the strongest evidence
for a rapid increase in the volume density of dusty, very luminous
galaxies in the distant Universe. Their density peaks at a redshift
$z \simeq 2.4$ at a value about 1000 
times greater than that at the present epoch. Thus we conclude that the
balance between the star-formation density in the most luminous
dust-obscured versus
unobscured galaxies has altered radically over the last 80\% of the
history of the Universe. 

\vfill\eject


\noindent
{\bf References.}

\refis{Barger98}
      Barger, A.\ J., Cowie, L.\ L.,  Sanders, D., Fulton, E., Taniguchi, Y., 
	Sato, Y., Kawara, K., Okuda, H.
       Submillimetre-wavelength detection of dusty star-forming galaxies at 
	high redshift.
      {\it Nature} {\bf 394,} 248--251 (1998). 

\refis{Barger99}
      Barger, A.\ J., Cowie, L.\ L.,  Smail, I., Ivison, R. J.,
      Blain, A. W., \& Kneib, J.-P. 
      Redshift distribution of the faint submillimeter galaxy population.
      {\it Astron. J.} {\bf 117,} 2656--2665 (1999).

\refis{BCR}
      Barger, A.\ J., Cowie, L.\ L., Richards, E.\ A.
      Mapping the evolution of high-redshift dusty galaxies with
      submillimeter observations of a radio-selected sample.
      {\it Astron. J.} {\bf 119,} 2092--2109 (2000).  

\refis{Barger02}
      Barger, A.\ J., Cowie, L.\ L., Brandt, W.\ N., Capak, P., 
      Garmire, G.\ P., Hornschemier, A.\ E., Steffen, A.\ T., Wehner,
      E.\ H.
      X-ray, optical and infrared imaging and spectral properties of the 1-Ms  
      Chandra Deep Field North sources.
      {\it Astron. J.} {\bf 124}, 1839 (2002). 

\refis{BSIK}
      Blain, A.\ W., Smail, I., Ivison, R.\ J., Kneib, J.-P.
      The history of star-formation in dusty galaxies.
      {\it Mon. Not. R. Astron. Soc.} {\bf 302,} 632--648 (1999).

\refis{B02}
      Blain, A.\ W., Smail, I., Ivison, R.\ J., Kneib, J.-P., Frayer,
      D.\ T.
      Submillimeter galaxies.
      {\it Physics Reports} {\bf 369,} 111--196; also available as preprint 
      astro-ph/0202228 at http://www.arXiv.org (2002).

\refis{Boyle}
      Boyle, B.\ J., Shanks, T., Croom, S.\ M., Smith, R.\ J.,
      Miller, L., Loaring, N., Heymans, C.
      The 2dF QSO redshift survey -- I.\ The optical luminosity
      function of quasi-stellar objects.
      {\it Mon. Not. R. Astron. Soc.} {\bf 317,} 1014--1022 (2000).

\refis{Chapmanradio01}
      Chapman, S.\ C., Richards, E.\ A., Lewis, G.\ F., Wilson, G.,
      Barger, A.\ J.
      The nature of the bright submillimeter galaxy population.
      {\it Astrophys. J.} {\bf 548,} L147--L151 (2001).

\refis{Chapmanmega02}
      Chapman, S.\ C., Barger, A., Cowie, L., Borys, C., 
      Capak, P., Fomalont, E., Lewis, G., 
      Richards, E., Scott, D., Steffen, A., Wilson, G., Yun, M.,
      The properties of microjansky radio sources in the HDF-N, SSA13, and
      SSA22 Fields.
      {\it Astrophys. J.} {\bf 585}, 57 (2003).

\refis{Chapmanradio02}
      Chapman, S.\ C., Lewis, G.\ F., Scott, D., Borys, C., Richards, E.A.,
      Understanding the nature of the optically faint radio sources and their
	connection to the submillimeter population.
	{\it Astrophys. J.} {\bf 570,} 557--570 (2002). 

\refis{Chapman02d}
      Chapman, S.\ C., Helou, G., Lewis, G.\ F., Dale, D., 
      The bi-variate luminosity-color distribution of IRAS galaxies,
	and impications for the high redshift Universe.
      {\it Astrophys. J.} in the press;also available as
      preprint astro-ph/0301233 at http://www.arXiv.org (2002). 

\refis{ChapmanLens02}
      Chapman, S.\ C., Smail, I., Ivison, R., Blain, A., 
	The effect of lensing on the identification of SCUBA galaxies,
      {\it Mon. Not. R. Astron. Soc.} {\bf 335,} 17--21 (2002).

\refis{Condon}
      Condon, J.
      Radio emission from ordinary galaxies.
      {\it Annu. Rev. Astron. Astrophy.} {\bf 30,} 575--611 (1992). 

\refis{Cowie02}
      Cowie, L.\ L., Barger, A.\ J., Kneib, J.-P.
      Faint Submillimeter Counts from Deep 850 Micron Observations
      of the Lensing Clusters A370, A851, and A2390.
      {\it Astron. J.} {\bf 123,} 2197--2205 (2002).

\refis{Fixsen}
      Fixsen, D.\ J., Dwek, E., Mather, J.\ C., Bennett, C.\ L., 
      Shafer, R.\ A.
      The spectrum of the extragalactic far-infrared background
      from the COBE FIRAS observations.
      {\it Astrophys. J.} {\bf 508,} 123--128 (1998). 

\refis{Fomalont02}
	Fomalont, E.\ B., Kellermann, K.\ I., Partridge, R.\ B., 	
	Windhorst, R.\ A., Richards, E.\ A.
	The Microjansky Sky at 8.4 GHz.
	{\it Astron. J.} {\bf 123,} 2402--2416 (2002).

\refis{I98}
      Ivison, R.\ J., Smail, I.,
      Le Borgne, J.-F., Blain, A.\ W.,
      Kneib, J.-P., Bezecourt, J., Kerr, T.\ H., \& 
      Davies, J.\ K.
      A hyperluminous galaxy at $z$ = 2.8 found in a deep
      submillimetre survey.
      {\it Mon. Not. R. Astron. Soc.} {\bf 298}, 583--593 (1998).

\refis{I00}
      Ivison, R.\ J., Smail, I, Barger, A.\ J.,
      Kneib, J.-P., Blain, A.\ W.,
      Owen, F.\ N., Kerr, T.\ H., \& Cowie, L.\ L.
      The diversity of SCUBA-selected galaxies
      {\it Mon. Not. R. Astron. Soc.} {\bf 315,} 209--221 (2000).

\refis{I02}
      Ivison, R.\ J., Greve, T., Smail, I., Dunlop, J., Roche, N., Scott, S., 
      Page M., Stevens, J., Almaini, O., Blain, A., Willott, C., Fox, M., 
      Gilbank, D., Serjeant, S., \& Hughes D.
      Deep radio imaging of the SCUBA 8-mJy survey fields: submm source 
      identifications and redshift distribution.
      {\it Mon. Not. R. Astron. Soc.} {\bf 337}, 1 (2002).

\refis{Kennicutt}
      Kennicutt, R.\ C.
      Star formation in galaxies along the Hubble sequence.
      {\it Annu. Rev. Astron. Astrophys.} {\bf 36,} 189--232 (1998).

\refis{LRIS}
      Oke, J.\ B., Cohen, J.\ G., Carr, M., Cromer, J., Dingizian, A., 
      Harris, F.\ H., Labrecque, S., Lucinio, R., \& Schaal, W. 
      The Keck low-resolution imaging spectrometer.
      {\it Publ. Astron. Soc. Pac.} {\bf 107,} 375--385 (1995).

\refis{Pettini}
      Pettini, M., Shapley, A.\ E., Steidel, C.\ C., Cuby, J.-G., 
      Dickinson, M., Moorwood, A.\ F.\ M., Adelberger, K.\ L., \&
      Giavalisco, M.
      The rest-frame optical spectra of Lyman-break galaxies: star 
      formation, extinction, abundances and kinematics.
      {\it Astrophys. J.} {\bf 554,} 981--1000 (2001).

\refis{Richards00}
	Richards, E.\ A.
	The Nature of Radio Emission from Distant Galaxies: 
	The 1.4 GHZ Observations.
      {\it Astrophys. J.} {\bf 533,} 611--630 (2000).
	
\refis{SScott02}
	Scott, S.\ E., Fox, M.\ J., Dunlop, J.\ S.,
 	Serjeant, S., Peacock, J.\ A., Ivison, R.\ J.,
 	Oliver, S., Mann, R.\ G., Lawrence, A.,
 	Efstathiou, A., Rowan-Robinson, M.,
 	Hughes, D.\ H., Archibald, E.\ N., Blain, A., Longair, M.
 	The SCUBA 8-mJy survey - I. Submillimetre maps, 
	sources and number counts.
	{\it Mon. Not. R. Astron. Soc.} {\bf 331,} 817--837 (2002).

\refis{ESI}
      Sheinis, A.\ I., Bolte, M., Epps, H.\ W., Kibrick, R.\ I., Miller, 
      J.\ S., Radovan, M.\ V., Bigelow, B.\ C., \& Sutin B.\ M.
      ESI, a new Keck Observatory echellette spectrograph and imager.
      {\it Publ. Astron. Soc. Pac.} {\bf 114,} 851--865 (2002).

\refis{SIB}
      Smail, I., Ivison, R.\ J., \& Blain, A.\ W. 
      A deep submillimeter survey of lensing clusters: a new window on 
      galaxy formation and evolution.
      {\it Astrophys. J.} {\bf 490,} L5--L8 (1997).  

\refis{Smail02}
      Smail, I., Ivison R.\ J., Blain, A.\ W., \& Kneib, J.-P. 
      The nature of faint submillimetre-selected galaxies.
      {\it Mon. Not. R. Astron. Soc.} {\bf 331,} 495--520 (2002).

\refis{Smail03}
      Smail, I., et al., 
	A vigorous starburst in the SCUBA galaxy N2-850.4. 
	{\it Mon.\ Not.\ R.\ Astron.\ Soc.} (in the press); 
	preprint available at http://www.arXiv.org/astro-ph/0303128 (2003).

\refis{Steidel99}
      Steidel, C.\ C., Adelberger, K.\ L., Giavalisco, M., 
      Dickinson, M. \& Pettini, M.
      Lyman-break galaxies at $z$ $>$ 4 and the evolution of
      the ultraviolet luminosity density at high redshift.
      {\it Astrophys. J.} {\bf 519,} 1--17 (1999).   

\refis{Steidel02}
      Steidel, C., Hunt, M., Shapley, A., Adelberger, K., Pettini, M.,
      Dickinson, M. \& Giavalisco, M.
      The population of faint optically-selected AGN at $z\sim$ 3.
      {\it Astrophys. J.} in the press; also available as prerint 
      astro-ph/0205142 at http://www.arXiv.org (2002).

\endreferences

\bigskip
\bigskip

\noindent
{\bf Acknowledgments.}

\noindent
We are grateful to Chuck Steidel, Alice Shapley and Tim Heckman for 
insightful discussions.  SCC acknowledges support from NASA. IRS
acknowledges support from the Royal Society and a Philip Leverhulme
Prize Fellowship. NRAO is operated by Associated Universities Inc.,
under a cooperative agreement with the US National Science
Foundation. Data presented herein were obtained using the W.\ M.\ Keck
Observatory, which is operated as a scientific partnership among
Caltech, the University of California and NASA. The Observatory was
made possible by the generous financial support of the W.\ M.\ Keck
Foundation.

\bigskip
\noindent
The authors declare that they have no competing financial interests.

\bigskip 
\noindent
Correspondence and requests for materials should be addressed to Scott
Chapman (schapman@irastro.caltech.edu).

\vfill\eject

{\eightrm \noindent {\bf Table~1.} Positions of submm galaxies with
redshifts from our survey, along with their measured redshifts, submm
fluxes and optical magnitudes, infered dust temperatures $T_{\rm d}$
and bolometric luminosities $L_{\rm bol}$. $T_{\rm d}$ and $L_{\rm
bol}$ were calculated, based on the redshift, submm and radio flux
densities, assuming that the far-IR--radio correlation
applies. Sources are in order of decreasing flux density, as in Figure~1.
(SMM163704.3+410530 was previously published in\refto{ChapmanLens02}.)
Submm fluxes for sources in the Lockman (10\ hr RA) and
ElaisN2 (16\ hr RA) fields are from\refto{SScott02} while fluxes for
other sources were measured ourselves.
Six out of 10 galaxies have $I$-band magnitudes fainter than 23.5,
consistent with the optical flux distribution of blank-field submm
galaxies presented elsewhere,\refto{Chapmanmega02}
and confirming that our sample is representative of the population.
Galaxies with hotter inferred dust temperatures could have
their radio flux densities enhanced by contribution from an active
galactic nucleus.
Understanding the relationship between submm galaxies and other
high-redshift populations will be helped by
comparing the strength of the clustering, both within and
between the different classes. This will require
substantially larger samples than are currently available.
\settabs 7 \columns
\+{RA} &dec &redshift &S$_{850 \mu m}$ (mJy)  &$I$-mag &T$_d$ (K) &
L$_{\rm bol}$ (10$^{12}$\L$_\odot$) \cr\medskip
\+SMM105224.6&+572119 &2.429    &11.7 $\pm$3.4  &25.9& $21 \pm 5$ & 3.0$^{+2.0}_{-0.7}$ \cr 
\+SMM163704.3&+410530 &0.840    &11.2 $\pm$1.6  &21.5& $16^{+4.1}_{-2.2}$ & $0.58^{+.32}_{-.34}$ \cr 
\+SMM105230.6&+572212 &2.610    &11.0 $\pm$2.6  &23.0& $29 \pm 6$ & 6.9$^{+5.9}_{-3.3}$ \cr 
\+SMM163650.0&+405733 &2.376    &8.2 $\pm$1.7  &22.2& $54^{+11}_{-7}$ & 45$^{+53}_{-19}$ \cr 
\+SMM123600.2&+621047 &1.998    &7.9 $\pm$2.4  &23.6& $37^{+8}_{-4}$ & 10$^{+11}_{-4}$ \cr 
\+SMM131201.2&+424208 &3.419    &6.2 $\pm$1.2  &23.5& $48^{+10}_{-7}$ & 20$^{+18}_{-7}$ \cr 
\+SMM105207.7&+571907 &2.698    &6.2 $\pm$1.6  &26.0& $36^{+7}_{-4}$ & 7.8$^{+7.1}_{-2.3}$\cr 
\+SMM131212.7&+424423 &2.811    &5.6 $\pm$1.9  &26.4& $54^{+19}_{-10}$ & 30$^{+60}_{-16}$ \cr 
\+SMM163658.8&+405733 &1.189    &5.1 $\pm$1.4  &22.4& $25^{+5}_{-9}$ & $1.5 \pm 1.2$ \cr 
\+SMM105155.7&+572312 &3.699    &4.5 $\pm$1.3  &24.2& $59^{+26}_{-16}$ & 30$^{+82}_{-20}$ \cr 
}

\vfill\eject

\noindent{\bf Figure Legends.}

{\eightrm
\noindent
{\bf Figure 1.} The ten spectra from Table~1
with identified redshifts (smoothed to the
instrumental resolution) from the sample of 34 submm galaxies observed, 
spanning the $I$ = 22.2--26.4 range. 
(Note that SMM163704.3+410530 was previously published in\refto{ChapmanLens02}.)
Pilot observations,
using Keck's ESI\refto{ESI} spectrograph on 2001 July 16 in the 
ELAIS-N2\refto{SScott02} field yielded three spectra with strong emission lines.
Multi-object Keck observations of 31 submm galaxies using
LRIS\refto{LRIS} were made on 2002 March 18--19 in light cirrus
conditions (with good 0.8$''$ seeing) in the HDF \& SSA13
fields\refto{BCR,Chapmanmega02}, 
and the Lockman Hole and ELAIS-N2 fields\refto{SScott02}.  
All the spectra probe the observed wavelength range from 0.3--1$\,\mu$m.
Exposures times were 1.5--4.5 hr, split into 30-min integrations. Data
reduction followed standard techniques using custom IRAF scripts.
One-dimensional spectra were extracted and compared with template
spectra and emission-line catalogues to identify redshifts. 
All identifications are based on multiple lines, most prominently 
the Ly$\alpha$ line which varies
tremendously in both flux (ranging from 1 to 60$\,\mu$Jy) and
rest-frame equivalent width (3 to $>$100\AA). Weaker
stellar/interstellar/AGN features and/or continuum breaks were
detected in the spectra, strengthening the redshift
identifications. Two galaxies at the same redshifts were identified
in the vicinity of the 10 robust submm sources, as shown for the
companions SMMJ105155.7+572312-a (one of the few submm galaxies
showing Lyman-$\alpha$ in absorption) and SMMJ105155.7+572312-b.  The
highest S/N spectrum (SMMJ163650.0+405733 --\refto{Smail03}) includes features
indicative of both strong starburst activity (P-Cygni wind absorption
profiles, stellar/interstellar absorption lines) and a weak AGN, a mix
typical of hybrid starburst/Seyfert-2 galaxies.
None of our spectra have broad lines like
Type-I AGN. Narrow-line Type-II AGN with enhanced N$\,$V and/or C$\,$IV
emission are consistent with half of our spectra. Despite this
evidence for AGN in these galaxies, they appear to be relatively weak
in energetic terms as compared to those identified in X-ray or optical
surveys.

\medskip
\noindent

{\bf Figure 2.} The redshift histogram of our submm galaxy sample. We
describe the selection effects using an evolving model of the local 
far-IR luminosity function, in which the dusty 
galaxies are represented by a range of
template spectral energy distributions. The model has been tuned to
fit the statistical properties of the 
submm-radio galaxy population (surface density
on the sky and submm--radio colours).\refto{Chapman02d} 
As a range of spectral energy distributions are represented at 
each galaxy luminosity, a broader redshift
distribution results than if the luminosity was tied
one-to-one with the dust temperature.\refto{B02} We plot
the predicted redshift distributions for submm galaxies with
flux densities $S_{850 \mu m} >$ 5\ mJy (solid line) and radio sources with
30\ $\mu$Jy $<S_{1.4\ \rm GHz} <$ 500\ $\mu$Jy 
(dashed line). We expect to
miss sources lying between the submm and radio model curves due to our
requirement of a radio detection to pinpoint the submm source.  The
apparent deficit between model and data at $z$ $\sim$ 1.5 may be
indicative of the difficulty in obtaining spectra in the spectroscopic
desert that spans the range 
1.2 $< z <$ 1.8.  Note that the models are in good overall
agreement with the observed redshift distribution.

\vfill
\eject

\centerline{\hbox{\psfig{file=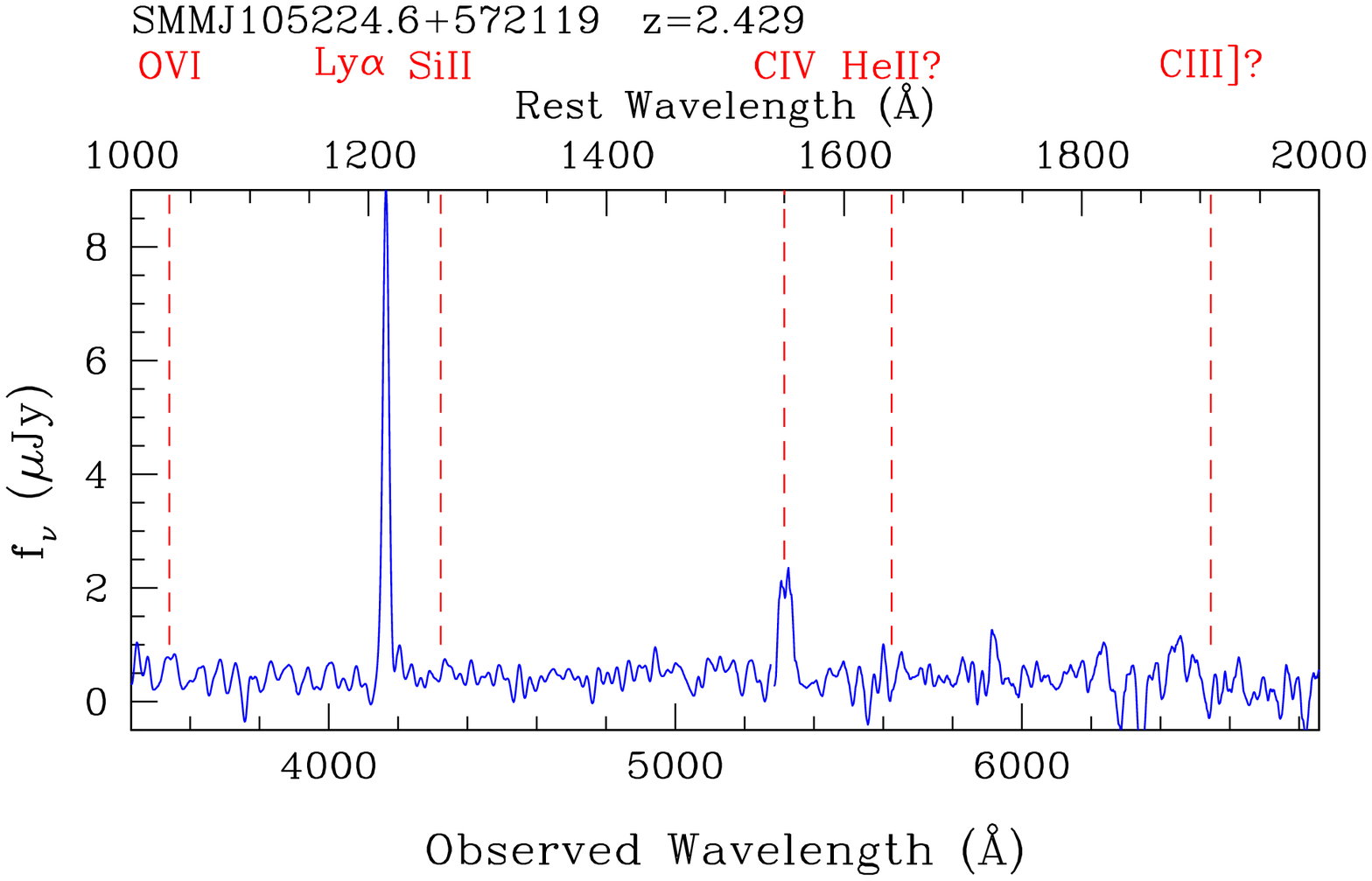,angle=0,width=3.5in}}}
\centerline{\hbox{\psfig{file=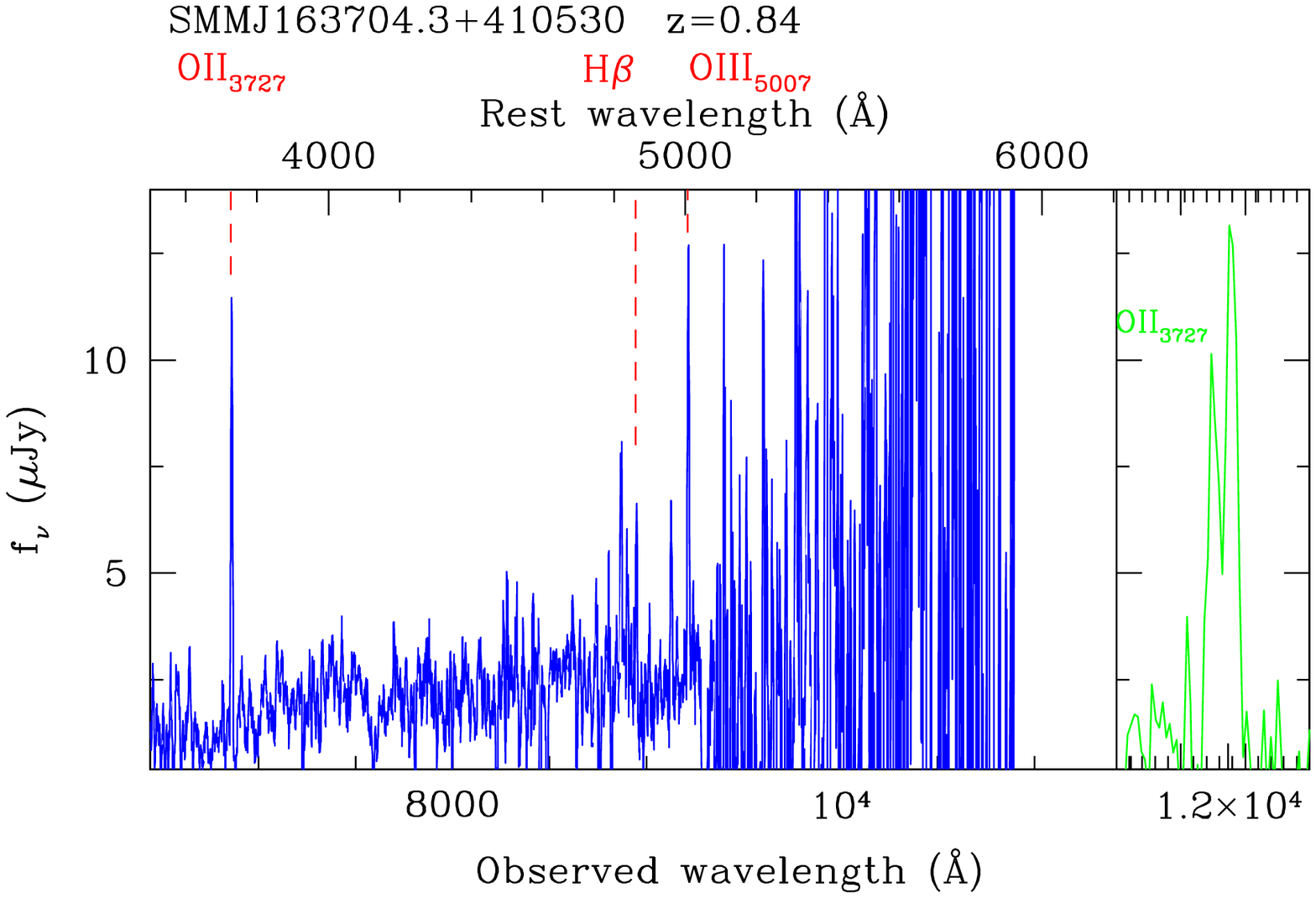,angle=0,width=3.5in}}}
\centerline{\hbox{\psfig{file=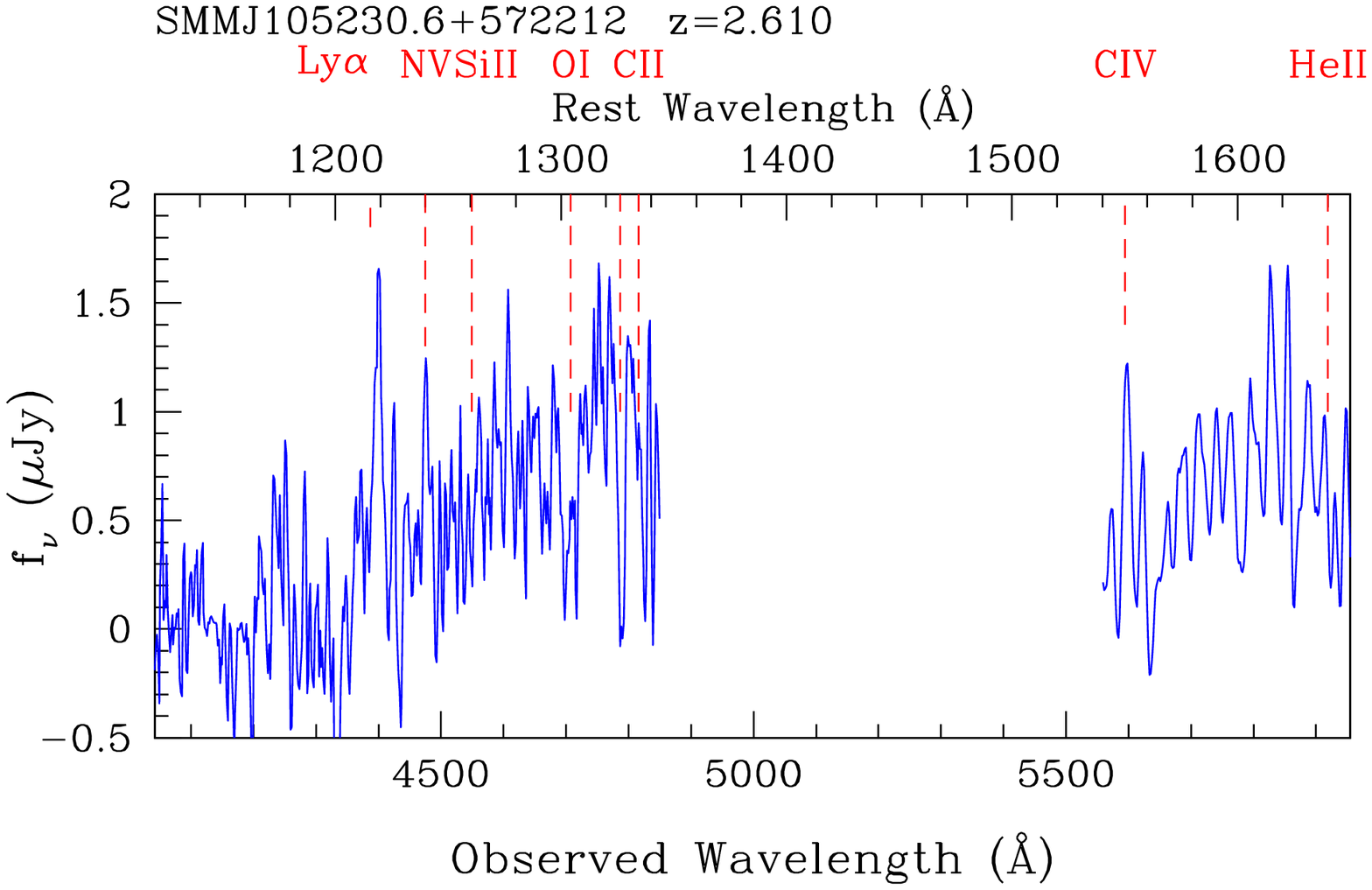,angle=0,width=3.5in}}}
\centerline{\hbox{\psfig{file=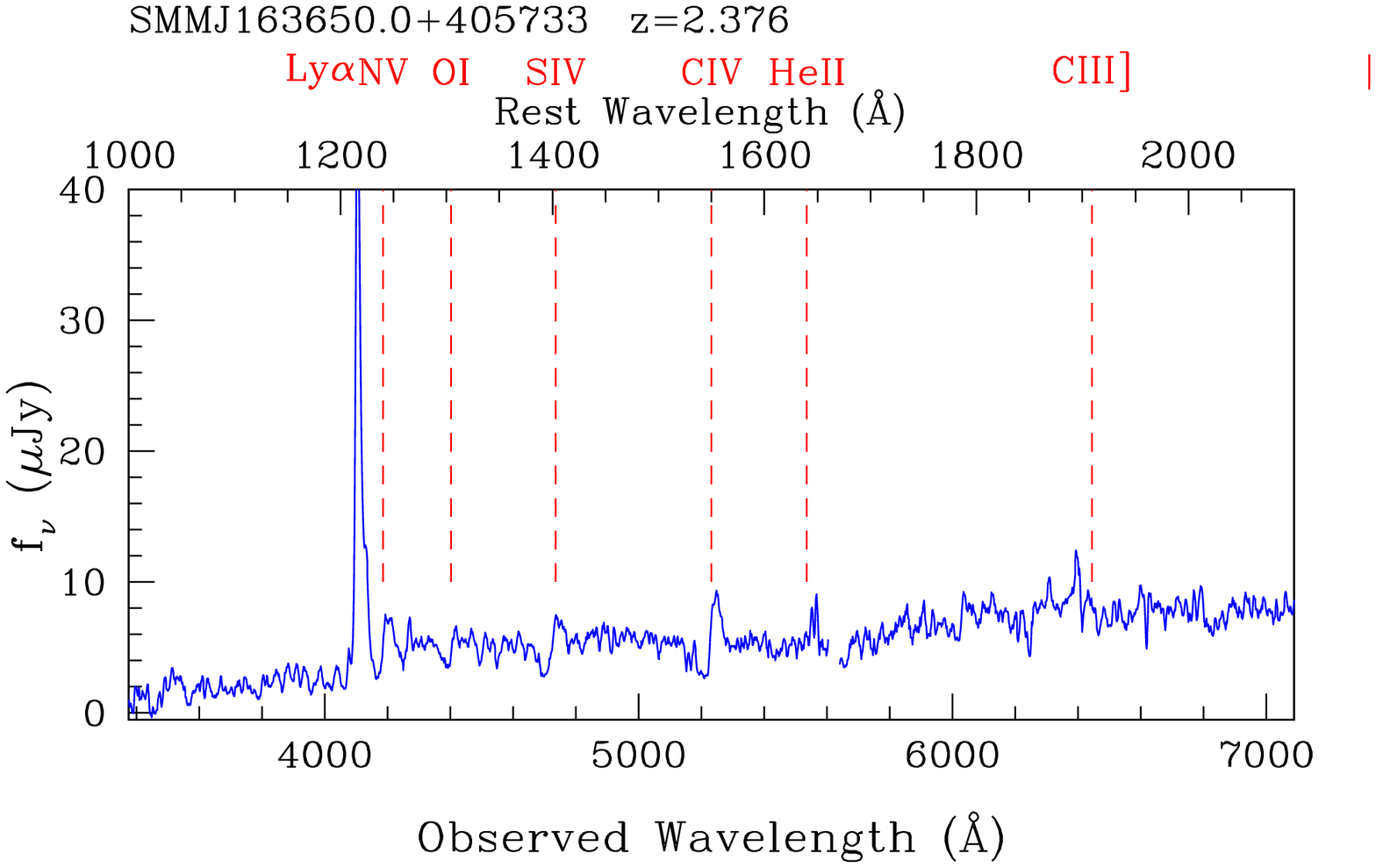,angle=0,width=3.5in}}}
\centerline{\hbox{\psfig{file=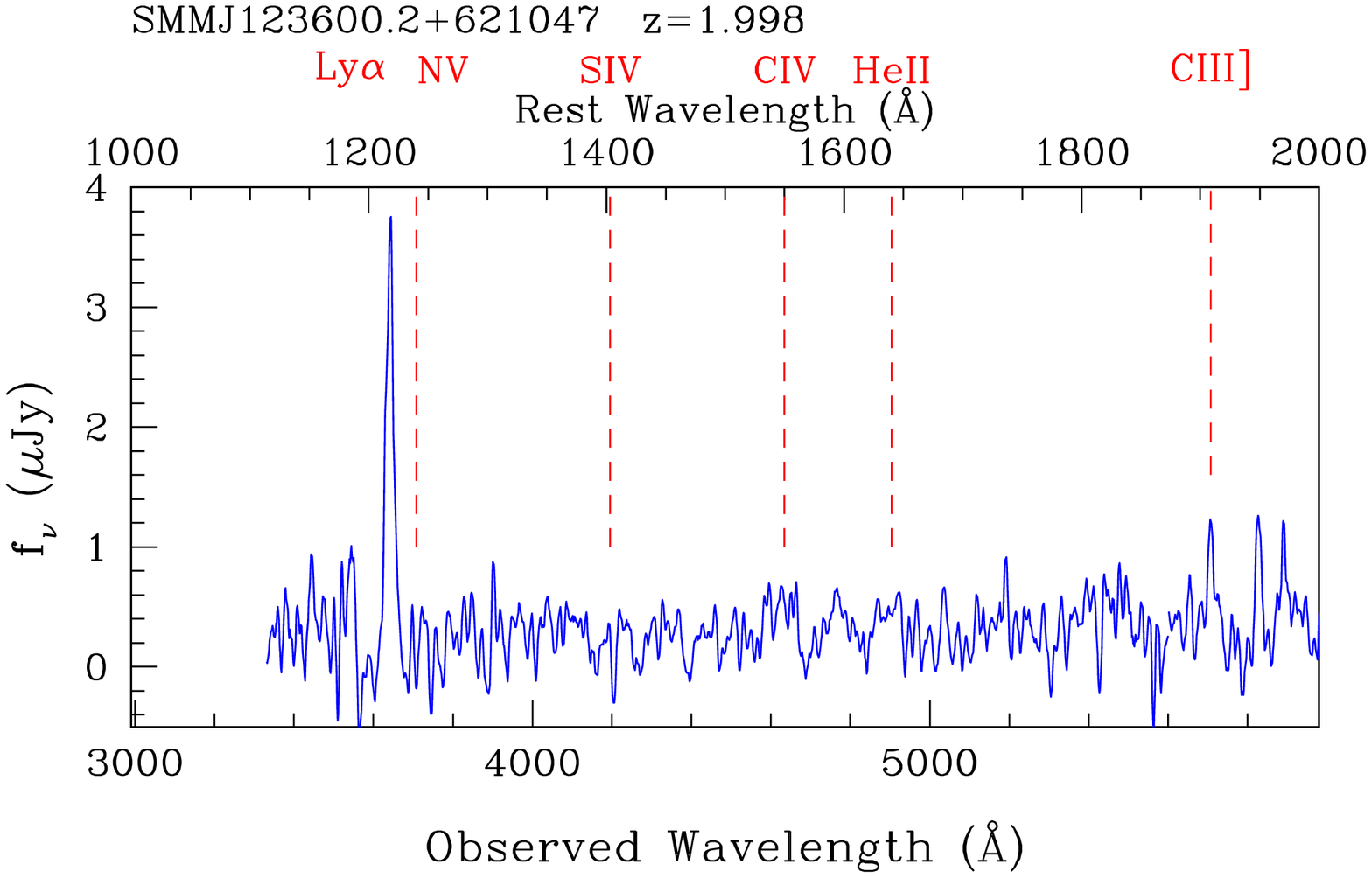,angle=0,width=3.5in}}}
\centerline{\hbox{\psfig{file=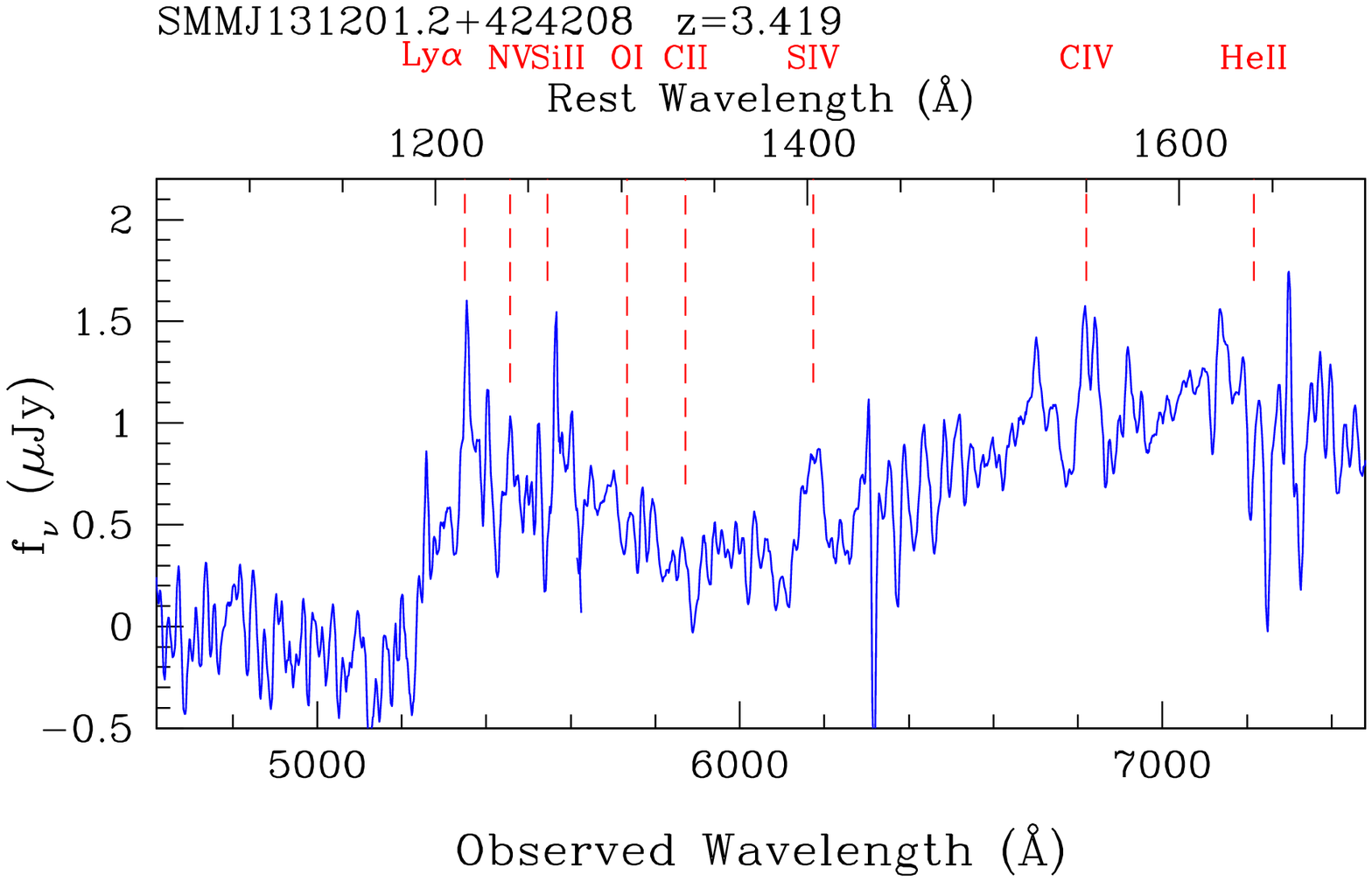,angle=0,width=3.5in}}}
\centerline{\hbox{\psfig{file=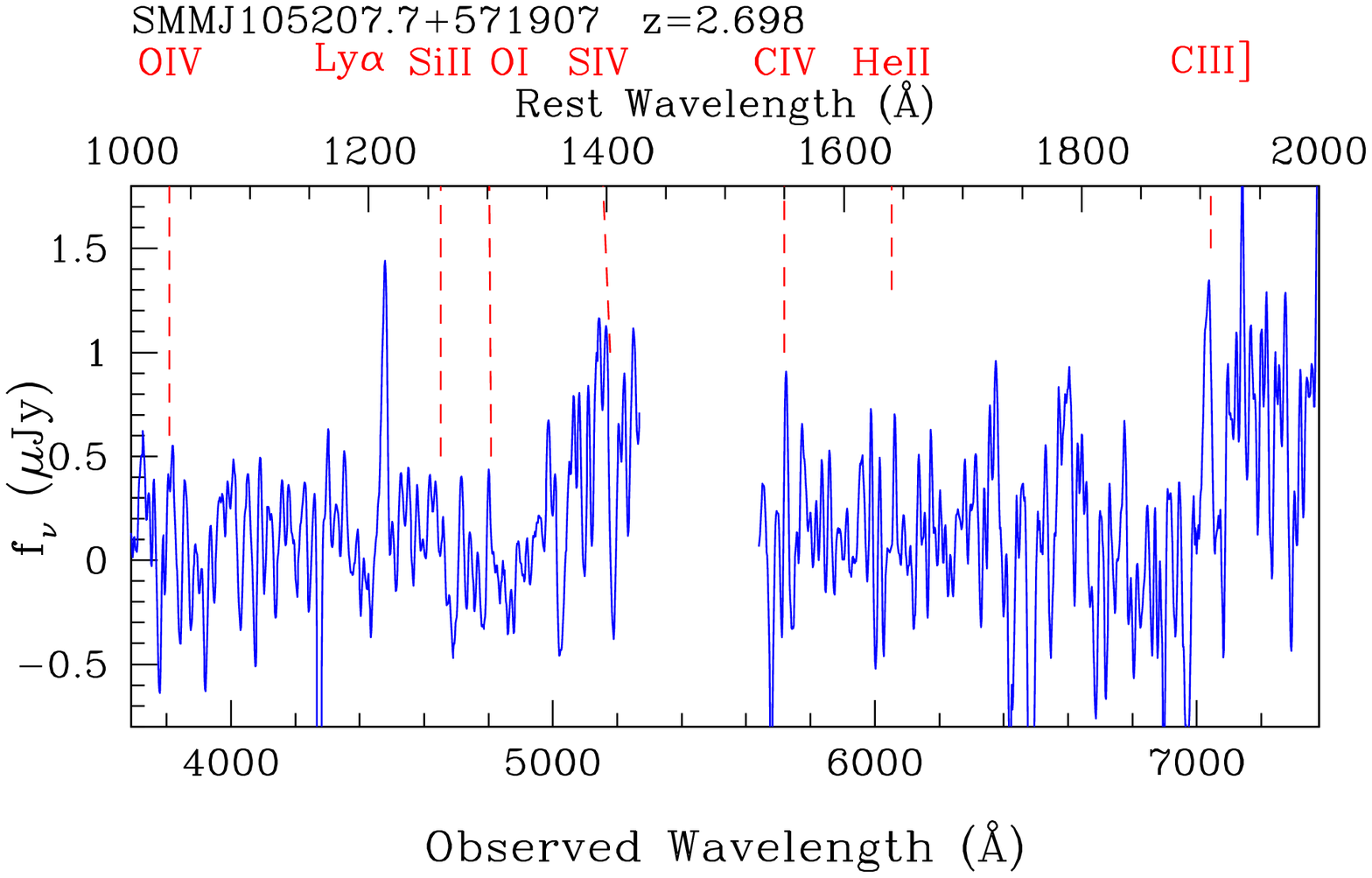,angle=0,width=3.5in}}}
\centerline{\hbox{\psfig{file=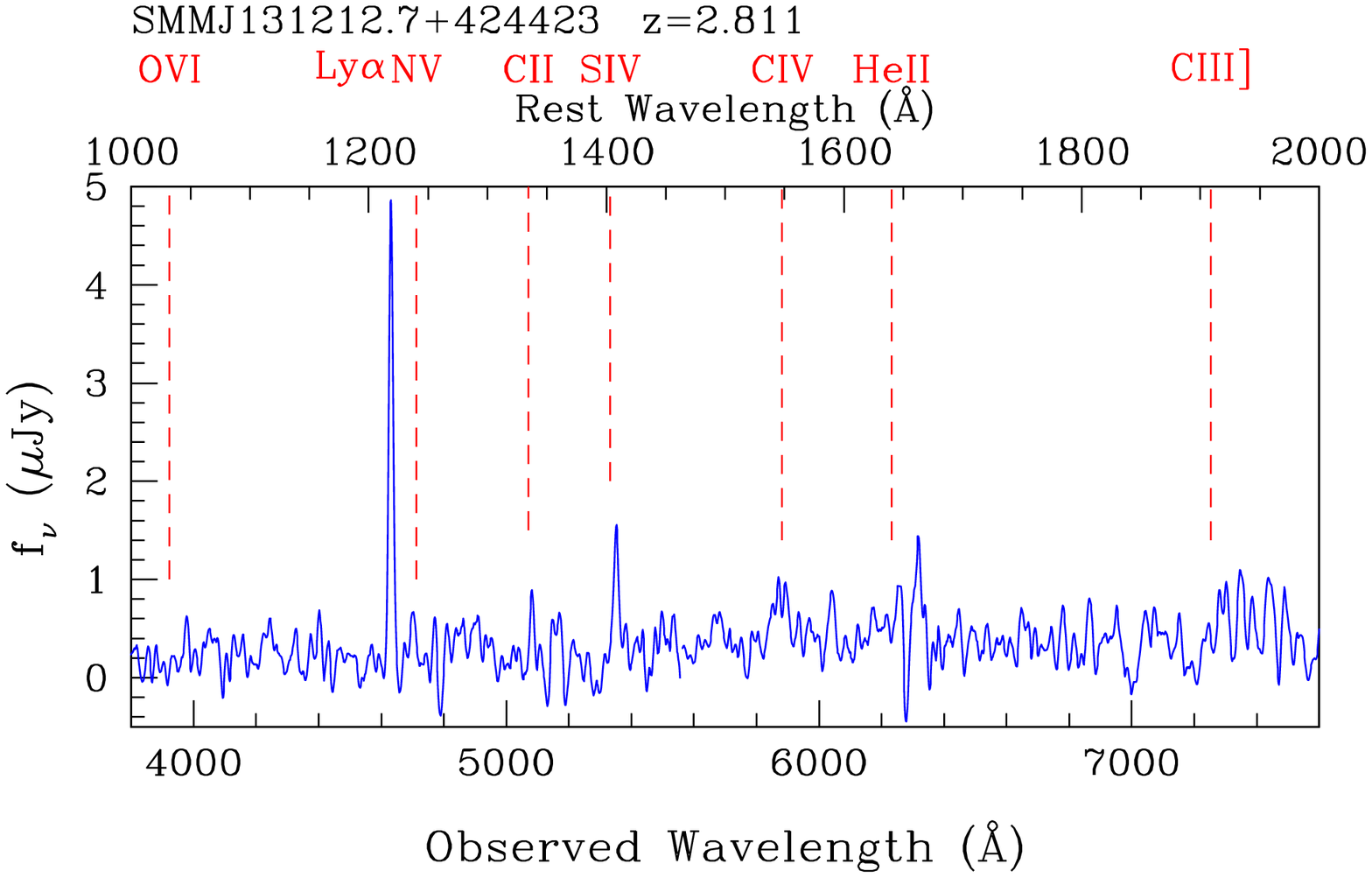,angle=0,width=3.5in}}}
\centerline{\hbox{\psfig{file=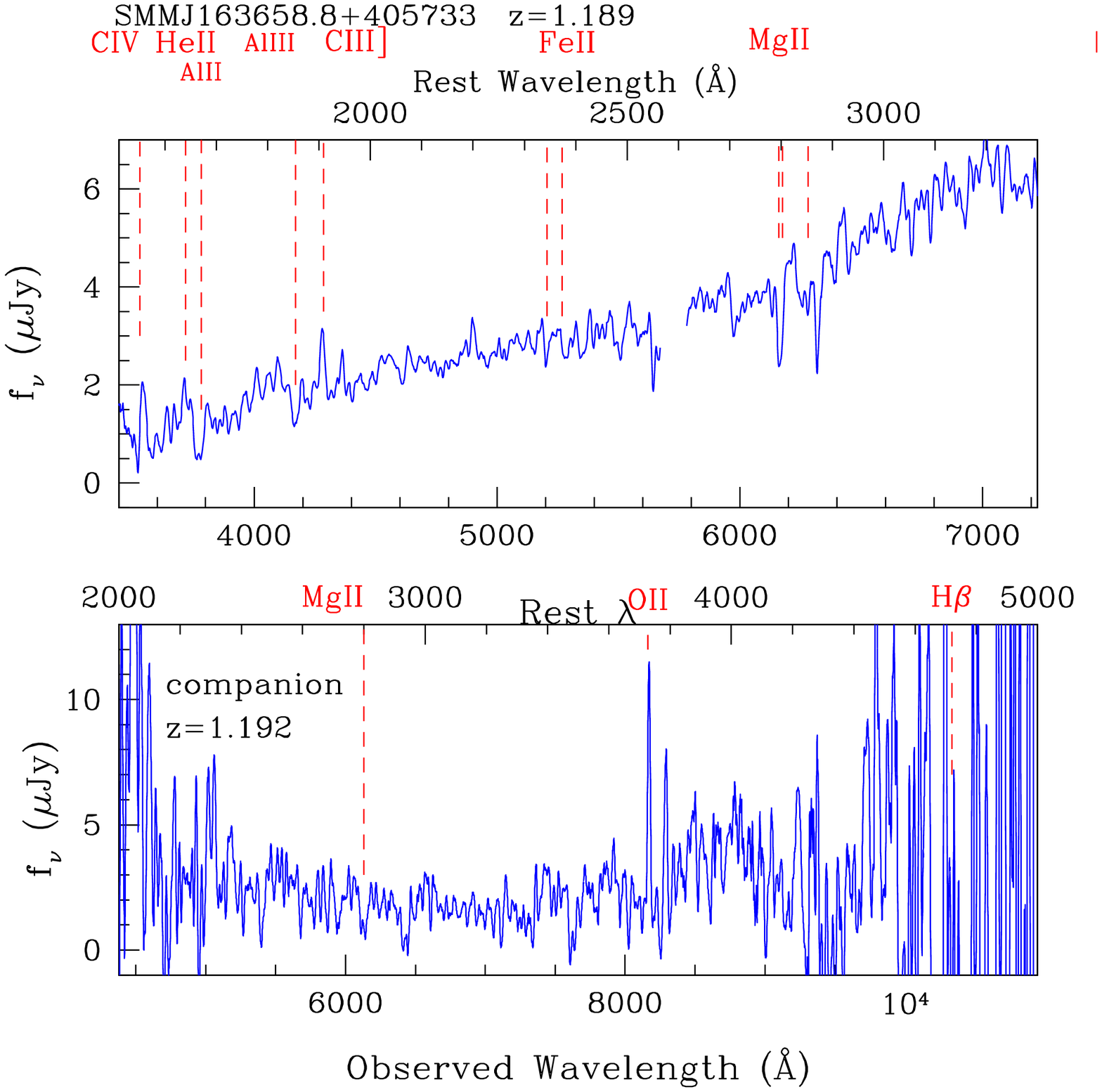,angle=0,width=3.5in}}}
\centerline{\hbox{\psfig{file=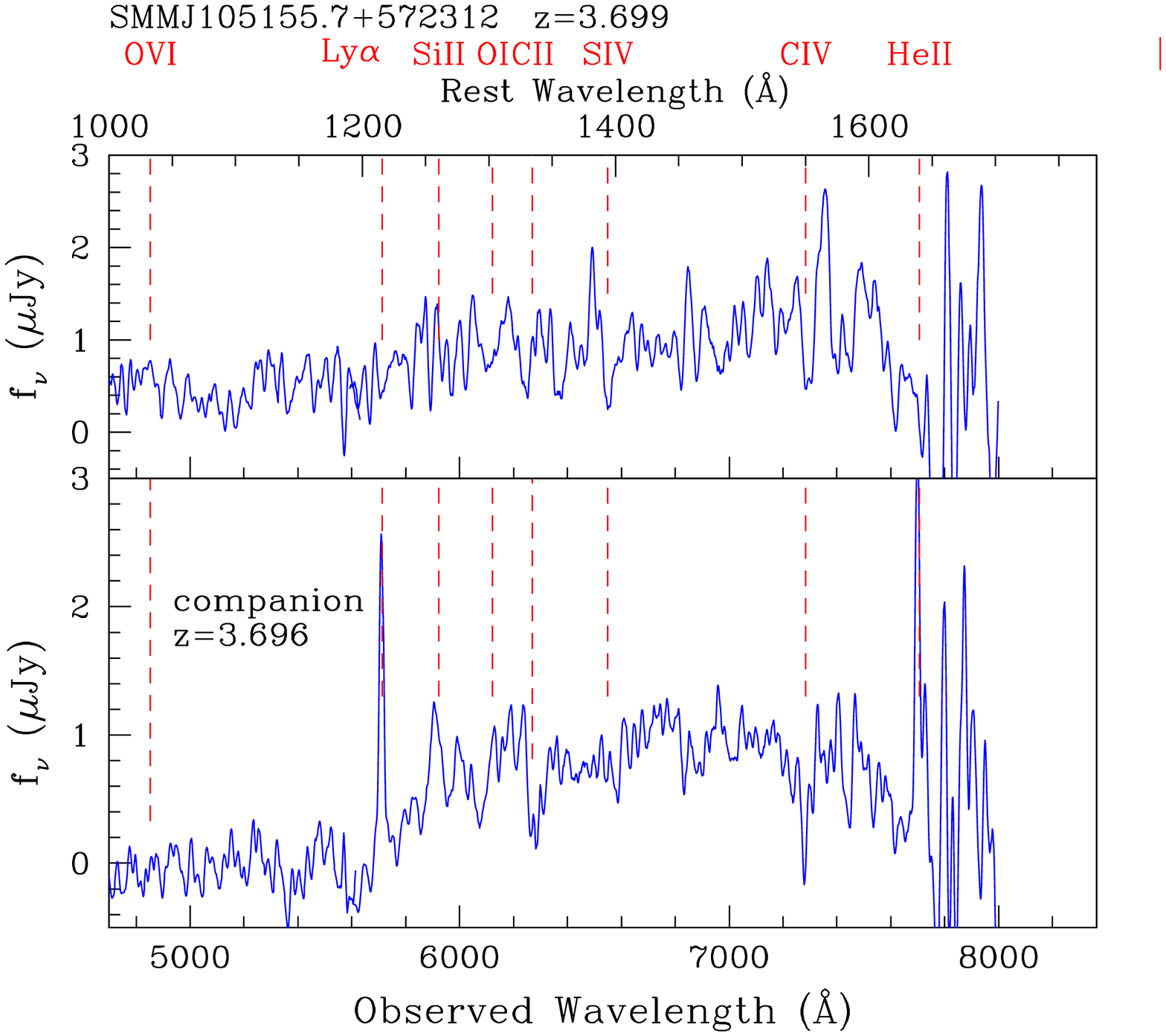,angle=0,width=3.5in}}}

\vfill
\eject

\centerline{\hbox{\psfig{file=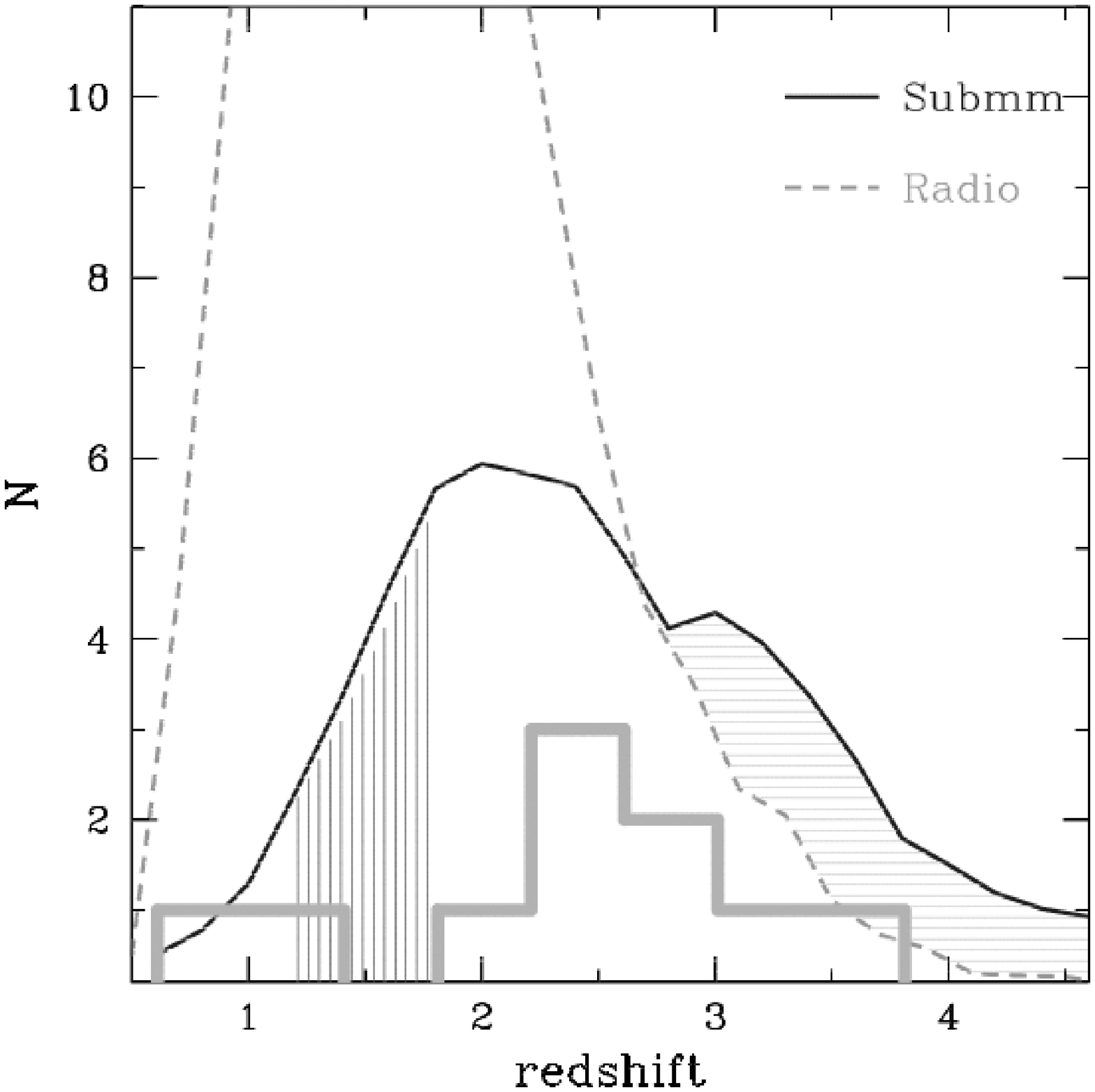,angle=0,width=5.5in}}}

\bye